\documentclass[]{aa}
\usepackage{graphicx}
 \usepackage{txfonts}
\begin{document}
%% Article title
%

  \title{Correlations between the peak flux density and the position angle of inner-jet in three blazars}
%% Running heads

%\shorttitle{Correlations between the peak flux density and the
%position angle of inner-jet in three blazars}

\author{X. Liu\inst{1,2}%\fnmsep\thanks{Corresponding author: e-mail: liux@xao.ac.cn}
           \and
            L.-G. Mi\inst{1,3}
           \and
           B.-R. Liu\inst{3,4}
           \and
           Q.-W. Li\inst{1,3}
}
 \offprints{X. Liu: liux@xao.ac.cn}

\institute{Xinjiang Astronomical Observatory, Chinese Academy of
Sciences, 150 Science 1-Street, Urumqi 830011, PR China \and Key
Laboratory of Radio Astronomy, Chinese Academy of Sciences,
Nanjing 210008, PR China \and Graduate University of the Chinese
Academy of Sciences, Beijing 100049, PR China \and College of
Physical Science and Technology, Guangxi University, Nanning
530004, Guangxi, PR China}

\date{Received / Accepted }

 \abstract
  {}
  {We aim to investigate the relation between the long-term flux
density and the position angle (PA) evolution of inner-jet in
blazars.}
   {We have carried out the elliptic Gaussian model-fit to
the `core' of 50 blazars from 15 GHz VLBA data, and analyzed the
variability properties of three blazars from the model-fit
results.}
  {Diverse correlations between the long-term peak flux
density and the PA evolution of the major axis of the `core' have
been found in $\sim$ 20\% of the 50 sources. Of them, three
typical blazars have been analyzed, which also show quasi-periodic
flux variations of a few years (T). The correlation between the
peak flux density and the PA of inner-jet is positive for
S5~0716+714, and negative for S4~1807+698. The two sources cannot
be explained with the ballistic jet models, the non-ballistic
models have been analyzed to explain the two sub-luminal blazars.
A correlation between the peak flux density and the PA (with a T/4
time lag) of inner-jet is found in [HB89]~1823+568, this
correlation can be explained with a ballistic precession jet
model. All the explanations are based mainly on the geometric
beaming effect; physical flux density variations from the jet base
would be considered for more complicated situations in future,
which could account for the no or less significance of the
correlation between the peak flux density and the PA of inner-jet
in the majority blazars of our sample.}

\keywords{BL Lacertae objects: individual: S5 0716+714, 3C~371,
4C~+56.27 -- radio continuum: galaxies -- galaxies: jets}
  % \keywords{BL Lacertae objects: individual: S5~0716+714 -- radio continuum: galaxies -- scattering}
   %\titlerunning{}
   \maketitle

\section{Introduction}

Blazars are thought of the extreme of active galactic nuclei
(AGNs), which consist of flat-spectrum radio quasars and BL Lac
objects. They are variable in almost whole electromagnetic
spectrum from radio to gamma-ray. The extreme properties of
blazars are mainly attributed to their jets which are pointing
closely to our line of sight, according to the unified model of
AGN (Urry \& Padovani 1995). With very long baseline
interferometry (VLBI) -- the highest angular resolution to date in
astronomy -- blazars can be resolved into a bright core and often
one-sided jet components in parsec scale or sub-parsec scale
(Lister et al. 2009a). In the VLBI scales, the blazar jets
frequently show fast outward motions, with an apparent advancing
speed of greater than the speed of light -- namely super-luminal
motion (Rees 1966). This is due to the so-called relativistic
Doppler boosting -- relativistic beaming effect (e.g. Cohen et al.
2007; Hovatta et al. 2009). Such high flux variability and VLBI
structure changes together provide us an effective means to study
the inner-jet property and probably the central engine's property
of the blazars in further.

Recently, Britzen et al. (2009) found there seems to be evidence
for an apparent stationarity of jet components (with regard to
their core separation) with time, in blazar S5~0716+714, while the
inner-jet components exhibit strong changes in their position
angles. This indicates that the jet components move non-radially
(or non-ballistic) with regard to their position angle. These
authors attributed such sub-luminal motions to a geometric origin,
although the geometric models were not well defined yet.
Furthermore, they found that the long-term flux density and the
position angle of the jet in 0716+714 show a significant positive
correlation. The similar phenomena of the apparent stationarity of
jet components with time have also been reported in BL Lac objects
S5~1803+784 and PKS~0735+178, by Britzen et al. (2010a, 2010b).
From CJF sample (Caltech-Jodrell Bank flat-spectrum sample of
radio loud active galaxies), Karouzos et al. (2012) investigated
the morphologies and pc-scale jet kinematics of more than 200
sources by using the so-called jet ridge-line method. They found
that about a half of the sample show jet widths $>10\degr$, with
BL Lac jet ridge lines showing significantly larger apparent
widths than both quasars and radio galaxies, indicating that the
BL Lacs have more significant position angle changes of jets in
their evolution.

From the study of AGN jet kinematics both in the 2 cm survey
(Kellermann et al. 2004) and in the radio reference frame image
database (RRFID), about one-third of well-measured component
trajectories are non-radial (Piner et al. 2007). In the MOJAVE
(monitoring of jets in active galactic nuclei with VLBA
experiments) sample, 166 out of the 526 robust components (32\%)
show non-radial motions, indicating that non-radial motion is a
common feature of the jet flow in blazars (Lister et al. 2009a,
2009b). Three of these non-radial components might be considered
inward at the 3$\sigma$ level (Lister et al. 2009b), this could be
due to a projection effect of curved jets. With the acceleration
measurements of the components of MOJAVE sources, Homan et al.
(2009) found the perpendicular accelerations are closely linked
with non-radial motions, and about half of the components show
`non-radial' motions; but the parallel accelerations are generally
larger than the perpendicular accelerations with respect to the
component velocities.

To further investigate the relation, which was first discovered in
0716+714 by Britzen et al. (2009), between the long-term flux
density and the position angle (PA) evolution of inner-jet in
blazars, we model-fitted the `core' components of 50
core-dominated balzars from the MOJAVE data released up to 2011
(Lister et al. 2009a). We aimed at finding more sources which
showing correlations between the long-term flux density and the PA
evolution of inner-jet. We also analyzed the inner-jet models to
explain the typical correlations, if any, in the model-fit
results.

\section{Model-fit result and analysis}
\begin{table*}

         \caption[]{The model-fit result of inner-jet in three blazars from the MOJAVE 15 GHz data,
         the result of correlation analysis, and the major period of peak flux variations derived from the PSD.}

         \tiny
         \renewcommand\baselinestretch{1.18}
       % \fns \tabcolsep 0.7mm

         \begin{tabular}{ccccccccccc}

\hline
  \hline
    \noalign{\smallskip}
    1&2 &3 &4 & 5&6 &7 & 8&9 &10  \\
    Source & Peak flux & Major axis& Minor axis& PA of M.A.& Dist. from ref. & PA of comp.  & Corr.2\&5 & Corr.5\&7 & T\\

  & [Jy/beam] & [mas] & [mas] &[deg]&  [mas] & [deg] &  & & [yr]\\

\hline

  \noalign{\smallskip}

%-----------------------------------------------------------------------------------------------------------------------
0716+714& 0.2258/1.4631/3.3709 & 0.11/0.21/0.37& 0.02/0.05/0.09 & 6.6/20.6/43.0  & 0/0.01/0.09  & -7.3/20.5/50.4 & 0.44(4.6E-4) & 0.62(2.1E-7) & $5.8\pm0.4$\\

%-----------------------------------------------------------------------------------------------------------------------
1807+698& 0.4826/0.6987/0.9561 &  0.50/0.68/0.91 & 0.09/0.11/0.23 & -110/-105/-99  & 0.04/0.07/0.14& -112/-104/-97 & -0.45(0.02) & 0.21(0.27) & $8.3\pm0.6$\\

%-----------------------------------------------------------------------------------------------------------------------
1823+568& 0.5676/1.1359/2.0515 & 0.31/0.40/0.70& 0.02/0.06/0.13 & -165/-161/-151 & 0.02/0.03/0.08 & -147/-146/-142 & 0.56(1.6E-5)$^{a}$ & 0.29(0.03)$^{a}$ & $7.0\pm0.4$\\

%-------------------------------------------------------------------------------------------------------------------------
           \noalign{\smallskip}
            \hline
           \end{tabular}{}
\tablefoot{$^{a}$Epoch of the PA of the major axis in 1823+568
data has been shifted forward by T/4 in the correlation analysis.}
       \label{tab1}
   \end{table*}

\subsection{Model-fit to source `core'}

We selected 50 core-dominated blazars showing one-sided VLBI
core-jet and they have been observed with the very long baseline
array (VLBA) at 15 GHz in more than 10 years, from the MOJAVE
database (Lister et al.
2009a\footnote{http://www.physics.purdue.edu/MOJAVE/}). We
model-fitted the `core' component of each source, with an elliptic
Gaussian model in the AIPS task `JMFIT', to get the peak flux
density and the PA of the major axis of the Gaussian component,
etc. We consider that the `core' in the 15 GHz VLBA images of
these core-dominated blazars is the `inner-jet' rather than the
true core. Therefore, the inner-jet can be modeled with an
elliptic Gaussian component which its major axis is along with the
inner-jet orientation or the inner-jet ridge-line on average, thus
reflecting the inner-jet PA.

We are interested in the long-term flux density and the PA
evolution of the `core' of blazars, 15 GHz MOJAVE datasets are the
best because these observations have been spanned over 10 years.
With the calibrated visibility of the MOJAVE data we could do the
model-fit to whole source structure, however, for the 50 sources
with more than 10 epochs of each source we couldn't complete the
multi-components model-fit to all the source structures, as
usually did with the modelfit program in the Difmap package. We
notice that it is not needed currently to model-fit whole source
structure, for that our purpose is studying on average the
innermost jet behavior of the sources. An advantage of our
approach is that we could get ride of an ambiguity in identifying
VLBI components from epoch to epoch which have long been discussed
as a potential problem for multi-epoch VLBI observations (see
discussions in e.g., Vermeulen et al. 2003; Piner et al. 2007;
Britzen et al. 2009). Our approach is similar to that the
model-fitting to the `core' of 250 flat spectrum radio sources by
Kovalev et al. (2005) with the visibility function.

The solution of the JMFIT in AIPS gives us the peak flux density
per beam, the X and Y position of a Gaussian component, the
de-convoluted major/minor axis and the PA of the major axis of the
Gaussian component. The internal error in the JMFIT is very small.
We have done twice the JMFIT for each dataset in order to get a
difference as the error of the solution. The final error of each
fitting parameter we adopted, is the larger one between the
internal fitting error and the difference of the twice fittings.

We have obtained the model-fit results of more than 500 epochs, we
here report only for three sources that have shown typical
correlations between the long-term flux density and the PA of the
major axis of `core'. We present the model-fit results for the
three sources: namely S5~0716+714, S4~1807+698, and
[HB89]~1823+568 in Table~\ref{tab1}. The columns give; (1) the
source name; (2),(3),(4),(5),(6),(7) the minimum /median /maximum
value of peak flux density per beam of elliptic Gaussian
component, major axis of the component, minor axis of the
component, position angle of the major axis, component position
shift from the reference point [0,0], position angle of the
component position shift, respectively; (8) linear Pearson
correlation coefficient (and significance) between the peak flux
density and the position angle of the major axis; (9) linear
Pearson correlation coefficient (and significance) between the
position angle of major axis and the position angle of component
shift; (10) major period of the peak flux variations derived from
the power spectrum density analysis.

\begin{figure}
    \includegraphics[width=7.5cm]{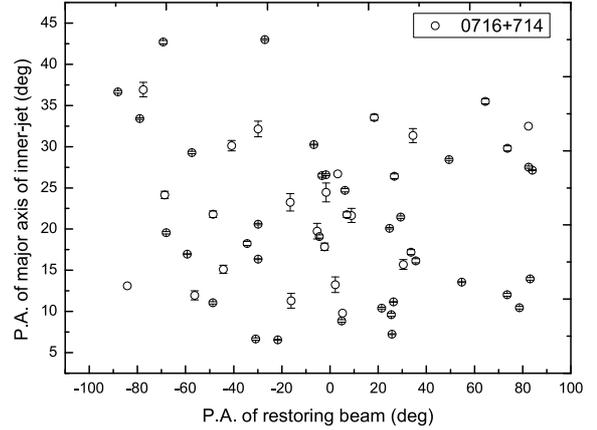}
    \caption{PA of the major axis of the 15 GHz VLBA `core' (inner-jet) versus PA of the restoring beam of the 15 GHz VLBA image for 0716+714.}
     \label{fig1}
   \end{figure}

It should be noted that there are no correlations between the
parameters of the de-convoluted elliptic Gaussian model from the
model-fit and the parameters of the restoring beam of image. For
instance, there is no any correlation between the PA of the major
axis of the 15 GHz VLBA `core' and the PA of the restoring beam of
the 15 GHz VLBA image for 0716+714, as shown in Fig.~\ref{fig1},
indicating that the model-fit result is related to the
source-intrinsic property rather than related to the restoring
beam of image.

\subsection{Correlation and PSD analysis}

The linear Pearson correlation is analyzed in the model-fit
parameters. A diversity of correlations between the long-term peak
flux density and the PA evolution of the major axis of the `core'
has been found in $\sim$ 20\% of 50 sources, including positive
correlation, anti-correlation, and correlation with some time
delay. There shows the positive, the negative and the time-delayed
correlation between the peak flux density and the PA of the major
axis for blazars 0716+714, 1807+698 and 1823+568 respectively in
Table~\ref{tab1} (also see, Fig.~\ref{fig2}, Fig.~\ref{fig3}, and
Fig.~\ref{fig4}). More discussions on the correlations are given
in the section 3. However, nearly 80\% of the 50 sources have
shown no or low significance of correlation between the long-term
peak flux and the PA evolution of major axis.

We also analyzed the possible periodicity of the long-term peak
flux variations, with the power spectrum density (PSD) method for
the three sources. All the three sources show a peak in the PSD
plot (Fig.~\ref{fig5}), the resulted major periodicity for
0716+714, 1807+698 and 1823+568, is respectively listed in
Table~\ref{tab1}.

\subsection{The three typical sources}

\subsubsection{S5~0716+714}

S5~0716+714 is a well known BL Lac object, with a redshift of 0.31
(Nilsson et al. 2008). The source is variable from radio to
gamma-ray (Wagner et al. 1996; Raiteri et al. 2003; Liu et al.
2012; Zhang 2010; Abdo et al. 2010), and it is also a TeV source
(Tavecchio et al. 2010). The VLA map at 20~cm shows kpc-scale jets
embedded in an oval cocoon (Antonucci et al. 1986). VLBI images
exhibit an extremely core-dominated jet pointing to the north and
misaligned with the VLA jet by $\sim90\degr$, e.g., at 5 and
8.4/22 GHz (Gabuzda et al. 1998). The 15 GHz MOJAVE images show a
bright compact core and a weak jet within 5 mas, with a jet
advancing speed of 10~c (Lister et al. 2009b). However, the
cross-epoch identification of components is not unique, from some
different data Britzen et al. (2009) identified the jet components
which are almost stationary respect to the core.

The correlation between the peak flux density and the PA of the
major axis of the `core' in 0716+714 is significant
(Table~\ref{tab1}). Though during the first part of the
double-peaks in 2009--2010, the PA is not significantly changed
but it does follow the trend of flux density variations (see
Fig.~\ref{fig2}). From long timescale radio data, a 5.7$\pm$0.5 yr
period is reported (Raiteri et al. 2003; Fan et al. 2007). From
our model-fit data at 15 GHz, the peak flux density does show a
peak in 2003--2004 and double peaks in 2009--2010
(Fig.~\ref{fig2}), with an overall periodic flux variations of
5.8$\pm$0.4 yr in the PSD plot (Fig.~\ref{fig5}, only data after
2001.0 were used in the PSD, because the VLBA data before 2001.0
are poor). There is also a possible period of 1.2$\pm$0.2 yr in
Fig.~\ref{fig5}.

\subsubsection{S4~1807+698}

The source 1807+698 (3C 371, z=0.051) is related to BL Lac objects
because of its optical variability and polarization in a large
elliptical galaxy whose emission is dominated by the flux from the
nucleus (Fan et al. 1999; Katajainen et al. 2000), and it is
detected in gamma-ray by Fermi LAT. VLA image at 1.36 GHz shows a
kpc scale halo and jets elongated in east-west (Cassaro et al.
1999). VLBI maps reveal a one-sided core-jet at PA
$\sim-100\degr$, e.g., at 5 GHz (G\'omez \& Marscher 2000). The
jets in the optical with HST and in the X-rays with Chandra have
been detected (Pesce et al. 2001), which are roughly aligned with
the parsec-scale radio jet. The 15 GHz MOJAVE result shows a
sub-luminal jet motion of 0.1~c (Lister et al. 2009b). A 43 GHz
VLBA image exhibits a jet flow assembling a wiggling pattern
(Lister 2001).

An anti-correlation between the peak flux density and the PA of
the major axis of inner-jet, with the coefficient of $-$0.45 at a
confidence level of 98\%, is found, and a flux variability period
of 8.3$\pm$0.6 yr is estimated with the PSD method (see,
Fig.~\ref{fig3} and Fig.~\ref{fig5}).

\subsubsection{[HB89]~1823+568}

The source 1823+568 (4C +56.27, z=0.664), is also a BL Lac object,
and it is detected in gamma-ray by Fermi LAT. It shows a halo and
one-sided kpc-scale jet to east (Kollgaard et al. 1992). VLA
observation at 5 GHz by O'Dea, Barvainis \& Challis (1986) shows a
curved jet extending southward from the core and turning to the
east. The object is highly variable and has a strong linear
polarization both in radio and optical bands (Lister 2001). VLBI
maps reveal a compact core and jets extending to the south, e.g.,
at 5 GHz (Fomalont et al. 2000). The 15 GHz MOJAVE result shows a
super-luminal jet motion of 20.8 c (Lister et al. 2009b).

Our model-fit result from the MOJAVE 15 GHz `core', suggests a
significant correlation between the peak flux density and the T/4
time-shift-forward PA of the major axis of inner-jet in
Table~\ref{tab1}. The variability period of T=7.0$\pm$0.4 yr is
estimated with the PSD (see, Fig.~\ref{fig4} and Fig.~\ref{fig5}).

\begin{figure}
     \includegraphics[width=7.5cm]{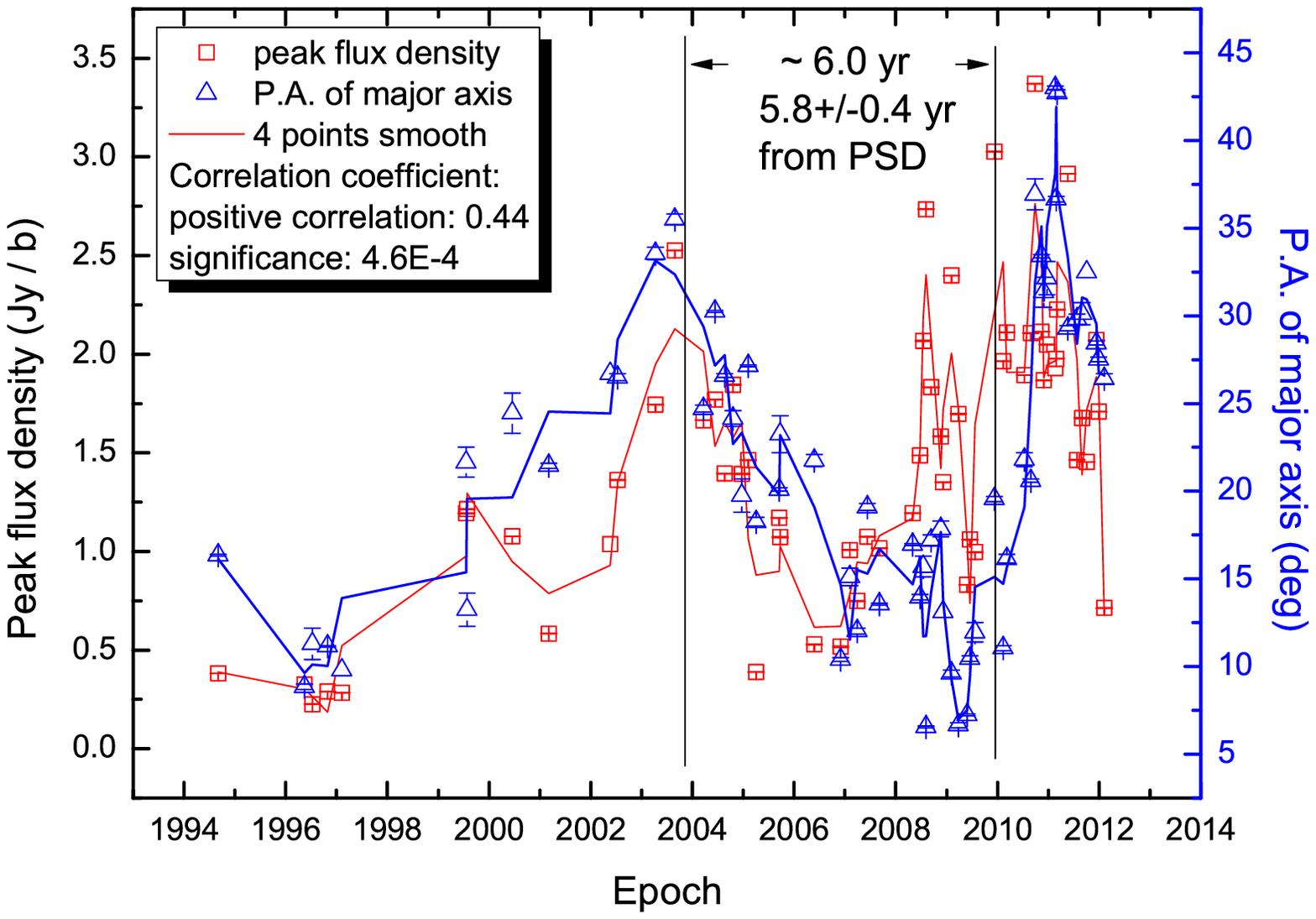}
     \caption{Peak flux density and position angle of the major axis of inner-jet in 0716+714 from the MOJAVE 15 GHz data versus epoch.
     Vertical lines show the estimated intervals from peak to peak of flux, and compared with the result from PSD analysis.}
      \label{fig2}
   \end{figure}

    \begin{figure}
     \includegraphics[width=7.5cm]{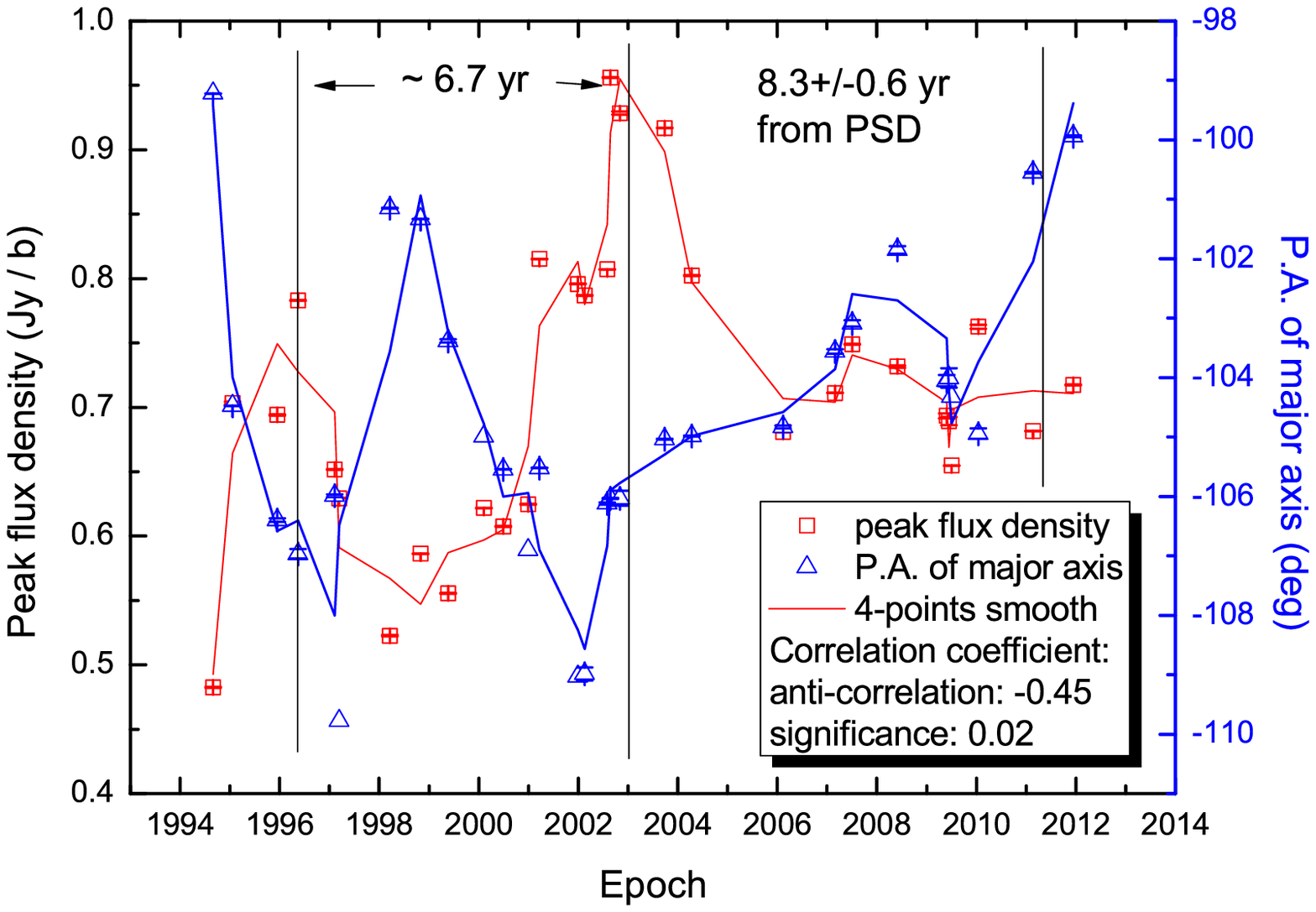}
     \caption{Peak flux density and position angle of the major axis of inner-jet in 1807+698 from the MOJAVE 15 GHz data versus epoch.
     Vertical lines show the estimated intervals from peak to peak of flux, and compared with the result from
     PSD analysis.}
      \label{fig3}
   \end{figure}

   \begin{figure}
     \includegraphics[width=7.5cm]{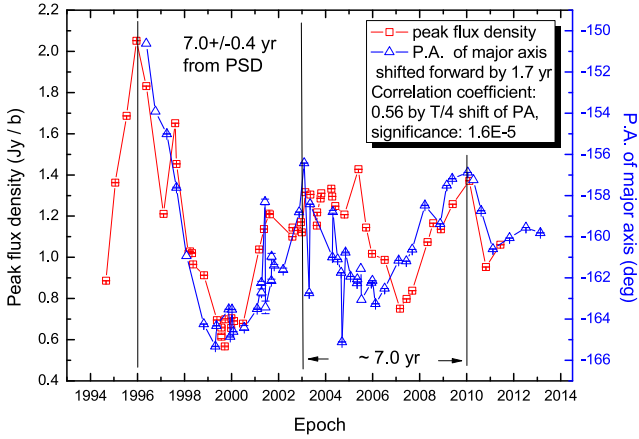}
     \caption{Peak flux density and position angle (shifted forward by 1.7 yr) of the major axis of inner-jet in 1823+568 from
     the MOJAVE 15 GHz data versus epoch. Vertical lines show the estimated intervals from peak to peak of flux, and compared with the result from
     PSD analysis.}
      \label{fig4}
   \end{figure}

 \begin{figure}
     \includegraphics[width=7.5cm]{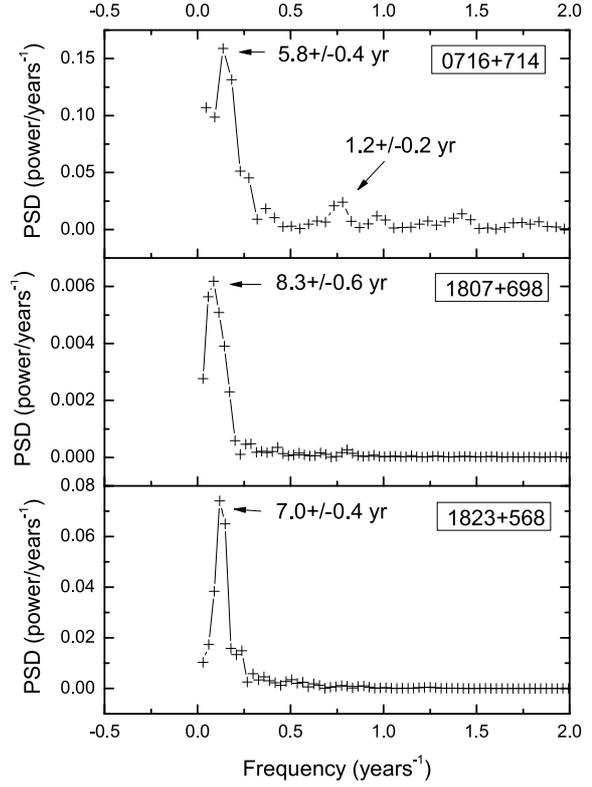}
     \caption{Power spectrum density of the peak flux density of 0716+714 (data with epochs $>$ 2001.0 were used, upper panel),
     1807+698 (middle panel) and 1823+568 (lower panel), respectively.}
      \label{fig5}
   \end{figure}

\begin{figure}
     \includegraphics[width=8cm]{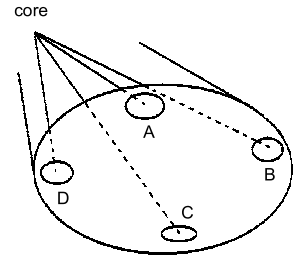}
     \caption{A schematic diagram of the ballistic jets with a precession
nozzle (i.e. the `B+P' model), with a counter-clockwise precession
`B$\rightarrow$A$\rightarrow$D$\rightarrow$C' for 1823+568, a jet
at the phase `A' is pointing the closest to our line of sight.}
      \label{fig6}
   \end{figure}

\begin{figure}
     \includegraphics[width=6.5cm]{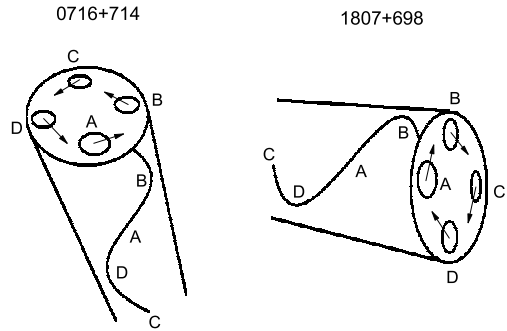}
     \caption{A schematic diagram of a non-ballistic jet with a counter-clockwise movement
`A$\rightarrow$B$\rightarrow$C$\rightarrow$D' for 0716+714 (left),
and a non-ballistic jet with a clockwise movement
`A$\rightarrow$B$\rightarrow$C$\rightarrow$D' for 1807+698
(right). In both cases, a jet at phase `D' is pointing the closest
to our line of sight.}
      \label{fig7}
 \end{figure}

\section{Inner-jet models and discussion}

We summarize our findings from the three BL Lac objects. There
exhibit diverse correlations between the peak flux density and the
PA of the major axis of `core' component in the three sources.
They also show quasi-periodic variations of the peak flux density,
around a few years.

To explain the correlations and periodicity mentioned above, we
have to find the models related to the inner-jet process in the
radio loud AGNs. In general, from the literature there are several
parameters/factors related to jet models, which should be
clarified and clearly defined, e.g., `ballistic' (we define as a
single jet component always moving along a straight line once
launched from the jet nozzle), `non-ballistic' (we define as a
single jet component moving in a curved way), `precession' (this
has been confused in the literature, we define it as a precession
`jet' in terms of a jet flow (by no means a single jet component),
in which every single jet component is ballistically launched from
a precession jet-nozzle), `non-precession' (we define as a jet
nozzle has no precession). There are four meaningful hybrids of
the parameters mentioned above, i.e., ballistic jets with a
non-precession nozzle (B+nP), ballistic jets with a precession
nozzle (B+P), non-ballistic jets with a non-precession nozzle
(nB+nP), non-ballistic jets with a precession nozzle (nB+P). The
quasi-periodic changing of the PA of inner-jet in the three
sources, could be explained with the hybrid models except the
first one (B+nP). A precession of the jet nozzle can be due to
e.g., double super-massive black holes or other reasons (see
Britzen et al. 2010a and references therein). For the
non-ballistic jet, models have been proposed e.g., Gong et al.
(2011), also see Lister et al. (2009b) and Britzen et al. (2009,
2010a, 2010b) for discussions. In these models, a non-ballistic
motion of jet is largely related to the jet interaction with
ambient matter. In the following, we will use the hybrid inner-jet
models we classified, for the three sources.

Firstly, we try to apply these models to 1823+568, after analysis
below, the B+P model (i.e. ballistic jets with a precession jet
nozzle) is applicable to this source, as shown in the schematic
diagram of Fig.~\ref{fig6}. We are nearly face-on the jet with the
estimated jet viewing angle of $\sim$8.4\degr (Savolainen et al.
2009), and the inner-jet PA is oscillating between $-$151\degr and
$-$165\degr (see Table~\ref{tab1} and Fig.~\ref{fig4}). In
Fig.~\ref{fig6}, our line of sight is close to the phase `A' of a
precession jet flow. At the phase `A', the relativistic beaming
effect is the strongest, so the flux density peaks; on the
contrast, the phase `C' of precession jet flow is far from our
line of sight, so the flux density is the lowest; at the phase `B'
or `D' the flux density has a middle value due to the beaming
effect. For the periodic peak flux variations (T$\sim$7.0 yr in
Table~\ref{tab1}), and the correlation between the peak flux
density and the PA (with a T/4 time-lag) of inner-jet, the
precession of the jet flow must be counter-clockwise, i.e. through
the phases
`B$\rightarrow$A$\rightarrow$D$\rightarrow$C$\rightarrow$B' in
Fig.~\ref{fig6}. Such that the PA peaks (a minus value) at the
phase `B' while the flux density is at a middle level; and the PA
has a middle value at the phase `A' while the flux density peaks;
and the PA has the smallest (minus) value at the phase `D' while
the flux density is at a middle level; and the PA has a middle
value at the phase `C' while the flux density is the lowest (also
see, Fig.~\ref{fig4}). So the correlation between the peak flux
density and the PA (with the T/4 time lag) of inner-jet can be
well explained. This `B+P' model seems to support the
super-luminal motions of 20.8~c (Lister et al. 2009b) or 9.4~c
(Savolainen et al. 2009) in this source, because the ballistic
motions have generally faster radial speed than that of the
non-ballistic motions. The origin of the periodic precession of
$\sim$7.0 yr in the jet nozzle still needs to be investigated in
future.

For the source either S5 0716+714 or 1807+698, the correlations
between the peak flux density and the PA of inner-jet cannot be
explained with a ballistic jet model. For 0716+714, the inner-jet
PA is oscillating between 6.6\degr and 43\degr (see
Table~\ref{tab1} and Fig.~\ref{fig2}). We are nearly face-on the
jet with the estimated jet viewing angle of $\sim$5.3\degr
(Savolainen et al. 2009). The significant positive correlation
(Table~\ref{tab1}) indicates that the flux density peaks while the
PA peaks (i.e. far from North) simultaneously (see
Fig.~\ref{fig2}), so it is impossible that the jet at the phase
`A' in the schematic diagram (Fig.~\ref{fig7} left) has a
middle-level of flux density according to the beaming effect
within a ballistic model. Non-ballistic jet models have to be
invoked, in which a single jet component is moving in a curved
way. In the non-ballistic scenario, regardless the jet nozzle is
in precession or not, the jet must have a non-radial motion. For
0716+714, in the schematic diagram Fig.~\ref{fig7} (left), for a
non-ballistic jet moving along the counter-clockwise helical
trajectory `A$\rightarrow$B$\rightarrow$C$\rightarrow$D', i.e. the
curved way through which the jet at the phase `D' is pointing the
closest to our line of sight, such that the flux density peaks
while the PA peaks too. The peak flux density has a middle value
while the PA has a middle value too at the phase `A', and so on,
the positive correlation between the peak flux density and the PA
at other phases `B', `C' can also be explained with the
counter-clockwise non-ballistic helical jet model (see Fig.2 and
Fig.~\ref{fig7} left). The non-ballistic jet model will lead to a
less outward motion speed than that of a ballistic jet model,
assuming the originally-launched jet speed is the same. This is
consisting with the sub-luminal motion or the `stationary
scenario' of 0716+714, as proposed by Britzen et al. (2009). In
general, a non-ballistic jet is likely to be formed, through the
shocked jet plasma working on the surrounding medium (Gong et al.
2011), and a helical like trajectory of the jet ridge-line would
be expected.

For 1807+698, the inner-jet PA is oscillating between $-$99\degr
and $-$110\degr (see Table~\ref{tab1} and Fig.~\ref{fig3}). We are
not well face-on the jet for the estimated jet viewing angle of
$\sim$45.3\degr (Savolainen et al. 2009). A clockwise
non-ballistic jet motion can explain the anti-correlation (i.e. a
T/2 time lag) between the peak flux density and the PA of
inner-jet, i.e. a jet is moving through
`A$\rightarrow$B$\rightarrow$C$\rightarrow$D' in Fig.~\ref{fig7}
right, the jet through at the `D' is pointing the closest to our
line of sight. Such that the flux density peaks due to beaming
effect at the `D' while the PA has a smallest (minus) value, and
at the phases `A' and `C', both the peak flux density and the PA
have middle values, and at the phase `B' the flux density is the
lowest while the PA has the largest (minus) value. The source jet
is sub-luminal (0.1~c, Lister et al. 2009b), which is the
non-ballistic jet model expected. The jet viewing angle of
1807+698 is relatively large (45.3\degr), so that the beaming
effect is reduced. This may have led to the relatively low
significance of the correlation between the peak flux density and
the PA of inner-jet of this source (see Table~\ref{tab1}), and the
lower significance of the estimated period of the flux variations.

Among the three sources, the periodicity of flux variability and
the correlation between the peak flux density and the T/4-laged PA
of inner-jet in 1823+568 is a purely geometric effect according to
the ballistic precession jet model `B+P'. The quasi-periodicity of
flux variability and the correlations between the peak flux
density and the PA of inner-jet in 0716+714 and 1807+698, as the
non-ballistic models considered above, are also mostly caused by
the geometric beaming effect. For the non-ballistic models in
particular, however, even if the precession of the jet nozzle is
invoked, the repeating periodic flux variability frequently found
in blazars still needs to be fully explained. A kind of physical
(non-geometric) outburst from the jet nozzle, or a physically
periodic flux variations from the jet base would be considered for
0716+714 and 1807+698 in the future, which may be related to
perturbations induced by the changes of the accretion mode or by a
periodic intervening of a companying object. Multi-bands study of
the three blazars on e.g., the SED, is also needed, like that
recently being investigated for balzars statistically (see, Meyer
et al. 2011; Lister et al. 2011).

\section{Summary}

We have carried out the elliptic Gaussian model-fit to the `core'
of 50 blazars. Diverse correlations between the long-term peak
flux density and the PA evolution of the major axis of the `core'
have been found in $\sim$ 20\% of the 50 sources. Of them, three
typical blazars have been analyzed, which also show quasi-periodic
flux variations of a few years (T). The correlation between the
peak flux density and the PA of inner-jet is positive for
0716+714, and negative for 1807+698. The two sources cannot be
explained with the ballistic jet models, the non-ballistic models
have been analyzed to explain the two sub-luminal blazars. A
correlation between the peak flux density and the PA (with a T/4
time lag) of inner-jet is found in 1823+568, this correlation can
be explained by a ballistic precession jet model. All the
explanations are based mainly on the geometric beaming effect;
physical flux density variations from the jet base would be
considered for more complicated situations in future, which could
account for the no or less significance of the correlation between
the peak flux density and the PA of inner-jet in the majority
blazars of our sample.

\begin{acknowledgements}

We thank the reviewer Junhui Fan and also Matthew L. Lister for
helpful comments and suggestions on the manuscript, which have
improved the paper. This research has made use of data from the
MOJAVE database that is maintained by the MOJAVE team (Lister et
al., 2009, AJ, 137, 3718). This research has made use of the
NASA/IPAC Extragalactic Database (NED) which is operated by the
Jet Propulsion Laboratory, California Institute of Technology,
under contract with the National Aeronautics and Space
Administration. The work is supported by the National Natural
Science Foundation of China (Grant No.11073036) and the 973
Program of China (2009CB824800).

\end{acknowledgements}

\end{document}